# In search of a structural model for a thiolate-protected Au$_{38}$ cluster

*De-en Jiang,*[*,†] *Weidong Luo,*[‡,§] *Murilo L. Tiago,*[‡] *and Sheng Dai*[†]

[†]Chemical Sciences Division and [‡]Materials Science and Technology Division, Oak Ridge National Laboratory, Oak Ridge, Tennessee 37831 and [§]Department of Physics and Astronomy, Vanderbilt University, Nashville, Tennessee 37235

jiangd@ornl.gov

In search of a structural model for Au$_{38}$(SCH$_3$)$_{24}$

*To whom correspondence should be addressed. E-mail: jiangd@ornl.gov. Phone: (865)574-5199. Fax: (865) 576-5235.

Abstract: The structure of thiolate-protected gold cluster $Au_{38}(SR)_{24}$ has not been determined experimentally and the best available signature is its measured optical spectrum. Using this signature and energetic stability as criteria and $SCH_3$ for SR, we compare four candidate structures: two from others and two we obtained. Our models are distinct from others' in that thiolate groups form monomers and dimers of the staple motif (a nearly linear RS-Au-SR bonding unit). We examine the energetics and electronic structures of the four structures with density functional theory (DFT) and compute their optical spectra with time-dependent DFT. We show that our dimer-dominated model is over 2.6 eV lower in energy than the two models from others and 1.3 eV lower than our previous monomer-dominated model. The dimer-dominated model also presents good agreement with experiment for optical absorption.

## 1. Introduction

A recent article in Chemical Reviews has nicely summarized the preparation, assembly, and applications of Au nanoparticles (AuNPs).[1] A major breakthrough in AuNP research has been the Schiffrin method[2] that for the first time made possible facile synthesis of heat- and air-stable AuNPs of several nanometers in size. This method utilizes thiolate groups to form a surface coating that protects AuNPs, based on the favorable chemical interaction between gold and sulfur. The resultant AuNPs can be easily handled and chemically manipulated, allowing a wide range of chemical transformations being performed with them.

The Schiffrin method and its alternatives (in which $AuCl_4^-$ is reduced and thiolate groups are used) usually produce thiolate-protected AuNPs of mixed sizes, which can be isolated chromatographically and then characterized by mass spectrometry. Whetten[3] and Tsukuda[4] and their respective coworkers have isolated a fraction of AuNPs that has been assigned a composition of $Au_{38}(SR)_{24}$.[5] The measured optical spectrum shows an optical band-edge energy of about 0.9 eV and a characteristic peak at 2 eV. However, the structure of $Au_{38}(SR)_{24}$ has not been determined. As Tsukuda and coworkers put it, "further effort is required to determine the geometric structures of the Au:SR clusters".[4]

First principles computations based on density functional theory (DFT) can greatly help find stable structures for thiolate-protected AuNPs and one key is to figure out the structure of the protective layer at the Au-thiolate interface. Using $SCH_3$ as the SR group, two models have been proposed for $Au_{38}(SCH_3)_{24}$ previously: one is based on cyclic Au-SR tetramers as the protective layer,[6] and the other has a mixture of Au-SR bonding motifs.[7] We have proposed a model for $Au_{38}(SCH_3)_{24}$ based on the staple motif (a linear or nearly linear RS-Au-SR bonding unit with the two S legs bonded to the surface Au atoms; see Scheme 1),[8] inspired by a recently reported single-crystal structure of $Au_{102}(SR)_{44}$.[9] Covered with six staple monomers and three dimers (see a dimer example in Scheme 1), our model is significantly lower in energy than the previous two.

After our work was published,[8] a single-crystal structure of $Au_{25}(SR)_{18}$ was reported[10] back-to-back with a successful DFT computational prediction.[11] Surprisingly enough, the obtained structure of $Au_{25}(SR)_{18}$ consists of a high-symmetry $Au_{13}$ core covered with six staple dimers and no staple monomers. In a sense, this new finding is not so surprising in that our previous DFT-based molecular dynamics simulations and structural optimization yielded dimer formation at the cluster surface.[8] What is surprising is that there is no staple monomer at all on the surface of $Au_{25}(SR)_{18}$. This new structure of $Au_{25}(SR)_{18}$ prompted us to ask the question whether there should be more dimers or only dimers on the surface of $Au_{38}(SR)_{24}$. To address this question, in the present work we seek new models for $Au_{38}(SR)_{24}$ by artificially creating more dimer units on the cluster surface and then compare energetics with previous models and optical absorption with experiment, to shed light on the likely structure for $Au_{38}(SR)_{24}$. We will show that a new dimer-dominated model indeed shows superior stability and good agreement with experiment for optical absorption.

## 2. Method

The Vienna Ab Initio Simulation Package (VASP)[12,13] was used to perform DFT calculations with planewave bases and periodic boundary conditions. We employed the generalized-gradient approximation (GGA) in the PBE form[14] for electron exchange and correlation and projector augmented wave (PAW) method for the electron-core interaction.[15] Scalar relativistic PAW potentials and a converged 450 eV kinetic-energy cutoff were used. We found that a 33% increase in the cutoff (to 600 eV) lowers the total energy for each $Au_{38}(SCH_3)_{24}$ configuration by ~0.20 eV but affects the relative energy between configurations only slightly (by ~0.05 eV), indicating that a cutoff energy of 450 eV is sufficient because we are more concerned with the relative energetics. $Au_{38}(SCH_3)_{24}$ clusters were placed in a cubic box of a=25 Å and only the Γ-point was used for the k-sampling. The force tolerance for structural optimization was set at 0.025 eV/Å. Electronic absorption spectra were obtained by using time-dependent DFT in the local, adiabatic approximation (TD-LDA), as implemented in the PARSEC code.[16,17] The spectrum is evaluated as a first-order perturbation in the electron density produced by an external laser source. We calculated the energy of optical excitations by diagonalizing an eigenvalue problem for the density matrix, following the energy-representation, real-space formulation proposed by Casida.[18] In order to simulate temperature and disorder effects, we applied a Gaussian convolution to the computed spectra, with a dispersion of 0.07 eV.

## 3. Results and discussion

We first compare three previous structural models for $Au_{38}(SCH_3)_{24}$ (see Scheme 2). Structure **1** was proposed by Häkkinen and coworkers.[6] This structure has an $Au_8$ cubic core, a middle shell of $Au_6$ with each of the six Au atoms sitting atop a face center of the cube, and an outer layer of six Au-SCH$_3$ cyclic tetramers. **1** exemplifies a proposal that Au-SR polymers protect AuNPs.[19,20] Structure **2** was obtained by Garzon et al.[7] and resembles **1**, but it has a disordered core and a slight lower energy (by 0.20 eV from our computation; see Table 1). The outer shell of **2** consists of a mixture of Au-SCH$_3$ bonding motifs from isolated SCH$_3$ groups to Au-SCH$_3$ trimers. Assuming that staple monomers are preferred bonding motifs on AuNPs, we obtained structure **3** by creating monomers on the $Au_{38}$ surface.[8] **3** also has a disordered core and its outer layer comprises six staple monomers and four dimers. More pronouncedly, **3** is 1.6 eV more stable than **1**.

Recently obtained crystal structure of $Au_{25}(SR)_{18}$ has an ordered $Au_{13}$ core and six staple dimers (and no monomers) on the cluster surface,[10] which confirmed a computational prediction for $Au_{25}(SCH_3)_{18}$.[11] The predicted model for $Au_{25}(SCH_3)_{18}$ is significantly lower in energy than previous models and shows good agreement with experiment for optical absorption,[11] indicating that thermodynamic stability and optical absorption can serve as good criteria for seeking candidate structures for thiolate-protected gold nanoclusters. Prompted by these new findings about $Au_{25}(SR)_{18}$, we improved upon structure **3** for $Au_{38}(SCH_3)_{24}$ by artificially creating more dimer units on the cluster surface. After optimization, we obtained several structures which are more stable than **3**. The most stable among the dimer-dominated structures is **4**, which like **3**, also has a disordered core but has six dimers and three monomers on the surface (see Scheme 2). **4**'s energy is about 1.3 eV lower than **3**, indicating that dimers are indeed preferred over monomers on $Au_{38}(SCH_3)_{24}$. However, further increase in the number of dimers led to structures less stable than **4**. This result indicates that the structure for $Au_{38}(SCH_3)_{24}$ is most likely to have a mixture of staple monomers and dimers but more of the latter.

Scheme 3 shows what the structure of **4** looks like from outside to inside. The whole cluster is shown in Scheme 3a. After removing the methyl groups, one can clearly see the staple monomers and dimers on the cluster surface in Scheme 3b (just a different angle of view from **4** in Scheme 2). Then we remove staple motifs (that is, S atoms and Au atoms between S atoms) from the cluster, and one is left with a disordered $Au_{23}$ core (Scheme 3c). Further removing some Au atoms with low coordination number from the core, one obtains an inner $Au_{12}$ core resembling a loose hcp packing. For more views of the structure, one can visualize **4** using the coordinates provide in the Supporting Information.

Is there a way to further lower **4**'s energy? Ref. 8 showed that DFT-based simulated annealing helped achieve a lower energy for staple-covered $Au_{38}(SCH_3)_{12}$. So we performed simulated annealing on **4**, but the resultant structure has a higher energy and shows formation of $(CH_3S-Au)x-SCH_3$ (x > 2) polymer on the cluster surface. There could be several reasons for this undesirable result: we might not run the molecular dynamics simulations long enough, or the temperature was not right. Because these simulated annealing runs are computationally expensive and do not guarantee a more stable configuration, we did not pursue it further for **4**. We found that the simulated annealing approach works better for lower coverages, but for full coverage the stapling method of doing educated guesses seems to work better.[8]

We examined the electronic structures of the four candidate structures for $Au_{38}(SCH_3)_{24}$. We found that **1** has two degenerate states at the Fermi level and the next level up is only 0.06 eV higher. A spin-polarized calculation showed that the triplet is slightly more stable (by ~0.03 eV) than the closed-shell singlet, indicating that **1** has radical character and is probably very reactive and unstable. **2**, **3**, and **4** all have a significant HOMO-LUMO gap, reminiscent of a stable, closed-shell molecule. **3** has the largest HOMO-LUMO gap at 0.78 eV.

Recently, Tsukuda and coworkers[4] measured optical absorbance of a relatively pure fraction of $Au_{38}(SR)_{24}$ down to 0.5 eV and obtained an optical band-edge energy (that is, the optical gap) of ~0.90 eV and a characteristic peak at 2.0 eV which was observed before.[3] We computed optical spectra of the four candidate structures using the TD-DFT method and show them in Figures 1-4 for structures **1**-**4**, respectively, together with the experimental spectrum.[21] Tsukuda and coworkers obtained the optical band-edge energy by extrapolating the first significant rise in absorbance and we followed their method in determining the optical band-edge energy for **1**-**4** (see insets in Figures 1-4). We note that there is uncertainty in determining this gap energy with the extrapolation method, because of the arbitrariness in choosing the region used for extrapolation. Moreover, the peak widths in our computed spectra depend on the Gaussian dispersion chosen, which also affects the extrapolated optical band-edge energy. Taken together, the uncertainty in the determined optical band-edge energies is estimated to be ~0.1 eV. Below we discuss the computed optical spectra in comparison with experiment.

Due to its high-symmetry core, **1**'s transitions for photon energy < 1.3 eV are forbidden, and the first major peak is located at 1.90 eV, followed by many other peaks between 2.0 and 3.7 eV. Our computed optical spectrum for **1** is similar to a previous one.[6] Using the rise leading to the peak at 1.9 eV, we obtained an optical band-edge energy of 1.7 eV (Figure 1 inset). Judging from the absorbance onset and profile, one can conclude that **1**'s optical spectrum is very different from the experiment.

**2**'s optical spectrum features a gap of 0.86 eV, the first major peak at 1.4 eV, and two other major ones at 1.9 and 2.2 eV. The gap agrees well with experiment, but which of the two peaks (1.9 and 2.2 eV) corresponds to the characteristic experimental peak at 2.0 eV is unclear. **3**'s optical spectrum has an absorbance onset of 0.80 eV, slightly lower than experiment, and a relatively smooth rise after 1.2 eV. The characteristic experimental peak at 2.0 eV is not visible in **3**' optical spectrum. **4**'s optical spectrum shows an optical band-edge energy of 0.94 eV, in good agreement with experiment. More importantly, 4 has an absorbance peak (though relatively flat) at 2.0 eV, corresponding well to the characteristic experimental peak. Moreover, **4**'s absorbance profile between 2.0 and 3.5 eV also matches well with experiment.

To understand deeper the electronic transition near the optical gap, we plot in Figure 5 the two involved orbitals with the leading contribution to the first measurable adsorption line in the computed optical absorption[22] for each of the four structures. One can see that the transition is not necessarily from HOMO to LUMO (only the case for **3**) and contributes in various degrees to the first measurable line (for example, 59% for **1** while 95% for **4**). Due to its high symmetry, **1**'s HOMO-6 is quite evenly distributed, mainly around the Au atoms within the cluster, and the LUMO shows lower symmetry and is more localized near the cube core. **2**'s HOMO shows localization around some S and Au atoms, and its LUMO+2 concentrates more around Au atoms in the core. **3**'s case is similar to **2**'s. **4**'s HOMO-3 shows significantly more localization around S and Au atoms in the staple motifs, and its LUMO also displays more distribution around Au atoms in the core.

Given the recent findings about $Au_{25}(SR)_{18}$ and our present new model (**4**) for $Au_{38}(SCH_3)_{24}$, it is worthwhile to discuss which Au-thiolate bonding motifs should be preferred to protect the gold cluster core. There are two different proposals regarding the structure of the thiolate protective layer at the AuNP-thiolate interface. Nobusada et al.[19,20,23] and Hakkinen et al.[6] proposed that the Au-thiolate polymer protects AuNPs. We proposed that the staple monomer dominates the AuNP-thiolate interface.[8] Our proposal was inspired by the recently obtained single crystal structure of $Au_{102}(SR)_{44}$ that shows 44 thiolate groups forming 19 staple monomers and two dimers on the cluster surface.[9] In a previous study,[8] we have shown how the staple motifs change the electronic structure of the underlying Au cluster and how they stabilize the cluster. One major role we found for the staple motif is that the two S legs of the staple motif help pin the surface Au atoms. This role is reduced for the dimer because the middle S is saturated, but the two end S atoms are still bonded to the surface Au atoms. For Au-SR polymers and especially cyclic polymers, the anchoring role of the S atoms would be significantly reduced. Hence, when the thiolate coverage is low and there are many surface Au atoms available, the staple monomer should be the preferred form. When the coverage is high and there are less surface Au atoms available, staple monomers tend to fuse together and form dimers. In fact, we observed the appearance of dimers with the thiolate coverage for $Au_{38}(SR)_x$ (x from 2 to 24).[8] The ratio of dimers over monomers is likely to depend on the thiolate coverage and the cluster size.

In the light of discussion above, one can consider $Au_{25}(SR)_{18}$ and $Au_{102}(SR)_{44}$ as two defining cases. For $Au_{25}(SR)_{18}$, the small size and the high coverage lead to all dimer units on the cluster surface;[10,11]

for $Au_{102}(SR)_{44}$, the relatively large size and the moderate coverage lead to a majority of monomers (19 monomers versus two dimers). $Au_{38}(SCH_3)_{24}$'s size is in between $Au_{25}(SR)_{18}$ and $Au_{102}(SR)_{44}$ but closer to $Au_{25}(SR)_{18}$, and its thiolate coverage is high, so it is reasonable that dimers dominate the surface as structure **4** shows. But we note that although **4**'s energy is significantly lower than previous models, its optical absorption does not match experiment perfectly. Better structures may exist and our models point towards a potential direction to unravel the true structure of $Au_{38}(SR)_{24}$.

Another concern has to do with the reliability of the DFT method for thiolate-protected gold nanoclusters. In the present work, we employed the PBE form of GGA for electron exchange and correlation in our calculations. Although PBE is a popular choice, we note that there exist many other forms of GGA and a different GGA functional may change the relative energetics among the four configurations of $Au_{38}(SCH_3)_{24}$. For example, Häkkinen et al.[6] showed that the energetic ordering between **1** and **2** is reversed when BLYP is used. So how can we be sure that **4** is indeed the preferred configuration? On the one hand, we can not be absolutely sure; on the other hand, we are fairly confident because (a) computed optical spectrum for **4** matches experiment well and (b) PBE has been used successfully to predict the structure of a similar system, $Au_{25}(SR)_{18}$.[11]

## 4. Summary and conclusions

We have compared four candidate structures for $Au_{38}(SCH_3)_{24}$: two from others, one from a previous study of ours, and a new, improved model. Using DFT at the GGA-PBE level, we obtained their energetics and electronic structures. The dimer-dominated model **4** is found to be most stable, followed by the staple monomer-dominated **3**. The Au-SCH$_3$ cyclic tretramer-covered model **1** and the structurally similar **2** of mixed surface Au-SCH$_3$ bonding motifs are close in energy but substantially less stable than **4** (by over 2.6 eV). **2**, **3**, and **4** have a significant HOMO-LUMO gap, while **1** shows open-shell character. Computed optical spectra show that **4** shows better agreement with experiment regarding the absorbance onset and the characteristic experimental peak at 2.0 eV. **2**'s optical spectrum also agrees well with experiment in terms of absorbance onset, although it shows two peaks around 2 eV. **3**'s optical absorption rise from 1.2 eV to 2.5 eV rather smoothly. **1**'s optical spectrum is very different from experiment and those of the other three. Based on the energetics, electronic structure, and optical absorption, we conclude that **4** is by far the best candidate structure for $Au_{38}(SCH_3)_{24}$.

**Acknowledgement.** This work was supported by Office of Basic Energy Sciences, U.S. Department of Energy under Contract No. DE-AC05-00OR22725 with UT-Battelle, LLC. The authors thank Prof. H. Häkkinen and Prof. I. L. Garzon for providing structural files for $Au_{38}(SCH_3)_{24}$. The authors are grateful to Prof. T. Tsukuda for providing experimental optical absorption spectra and helpful discussion. This research used resources of the National Energy Research Scientific Computing Center, which is supported by the Office of Science of the U.S. Department of Energy under Contract No. DE-AC02-05CH11231.

**Supporting Information Available:** Coordinates for structures in Scheme 2 (TXT). This information is available free of charge from the Internet at http://pubs.acs.org.

[21] The experimental spectrum was published in Ref. 4 and kindly provided by Prof. T. Tsukuda. Both computed and experimental spectra are normalized to strongest absorbance below or at 4.0 eV. Unlike in Ref. 4, we did not use the Jacobian relationship when converting the experimental absorbance from against wavelength (nm) to against photon energy (eV).

[22] Computed optical absorption consists of a series of lines at different photon energies with the line height being the oscillator strength. These lines can be broadened with a Gaussian dispersion of chosen width (in our case, 0.07 eV) to give a spectrum as shown in Figure 1. When examining orbitals involved in transitions, we look for the absorption line with considerable height and the lowest photon energy which usually corresponds to the optical gap, and determine which transition and its invoved orbtials contribute the most to the absorption line.

[23] Iwasa, T.; Nobusada, K. *J. Phys. Chem. C* **2007**, *111*, 45.

Table 1. Relative energetics, HOMO-LUMO gap, and optical gap (all in eV) for four candidate structures of $Au_{38}(SCH_3)_{24}$.

| Structure | 1 | 2 | 3 | 4 |
| --- | --- | --- | --- | --- |
| Relative Energy [a] | 0.0 | -0.24 | -1.59 | -2.86 |
| HOMO-LUMO GAP [a] | 0 | 0.34 | 0.78 | 0.39 |
| Optical Gap [b] | 1.69 | 0.86 | 0.80 | 0.94 |

[a] From present DFT-PBE results.

[b] From present TD-DFT spectra and determined by extrapolation (see insets in Figures 1-4).

Scheme 1. One staple motif (the monomer) on $Au_{38}(SCH_3)_2$ and one dimer motif on $Au_{38}(SCH_3)_3$. Au in yellow, S in blue, C in red and H in green.

Scheme 2. Four candidate structures for $Au_{38}(SCH_3)_{24}$. Au in yellow, S in blue, C and H not shown.

Scheme 3. Structural details of **4**: (a) the whole cluster $Au_{38}(SCH_3)_{24}$; (b) removing $CH_3$ groups; (c) removing the staple monomers and dimers; (d) further removing Au atoms with low coordination number. Au in yellow, S in blue, C in red and H in green.

Figure 1. Computed optical absorption spectrum for structure **1**. Experimental spectrum from Ref. 4 is shown (dashed line) for comparison. Inset shows a zoom-in of the computed spectrum between 1.5 and 1.9 eV and the extrapolation line for optical band-edge energy. Both computed and experimental spectra are normalized to strongest absorbance at or below 4.0 eV. This normalization is applied in all subsequent figures.

Figure 2. Computed optical absorption spectrum for structure **2**. Experimental spectrum from Ref. 4 is shown (dashed line) for comparison. Inset shows a zoom-in of the computed spectrum between 0.4 and 1.5 eV and the extrapolation line for optical band-edge energy.

Figure 3. Computed optical absorption spectrum for structure **3**. Experimental spectrum from Ref. 4 is shown (dashed line) for comparison. Inset shows a zoom-in of the computed spectrum between 0.6 and 1.2 eV and the extrapolation line for optical band-edge energy.

Figure 4. Computed optical absorption spectrum for structure **4**. Experimental spectrum from Ref. 4 is shown (dashed line) for comparison. Inset shows a zoom-in of the computed spectrum between 0.5 and 1.6 eV and the extrapolation line for optical band-edge energy.

Figure 5. Orbitals involved in single-electron transition of leading contribution to the first measurable absorption line (near the optical gap) of the computed optical absorption for structures **1**, **2**, **3**, and **4**. Line position (in eV) and percentage of leading contribution are indicated. Isovalues for green and red isosurfaces are 0.32 and -0.32 $a_0^{-3/2}$, respectively. Au in yellow, S in blue, C and H not shown.

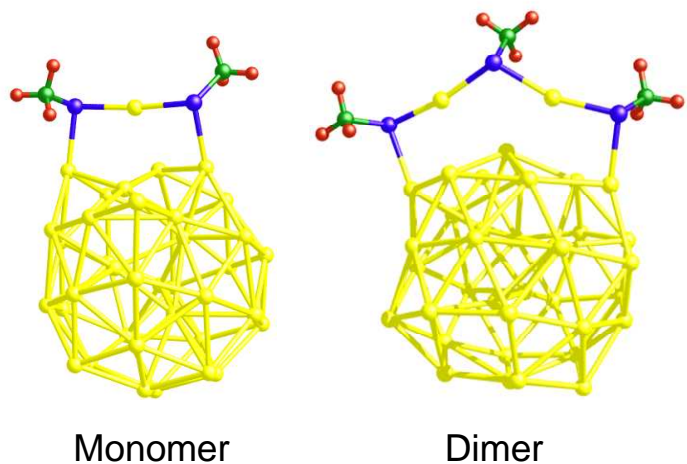

Scheme 1.

One staple motif (the monomer) on Au$_{38}$(SR)$_2$ and one dimer motif on Au$_{38}$(SR)$_3$. Au in yellow, S in blue, C in red and H in green.

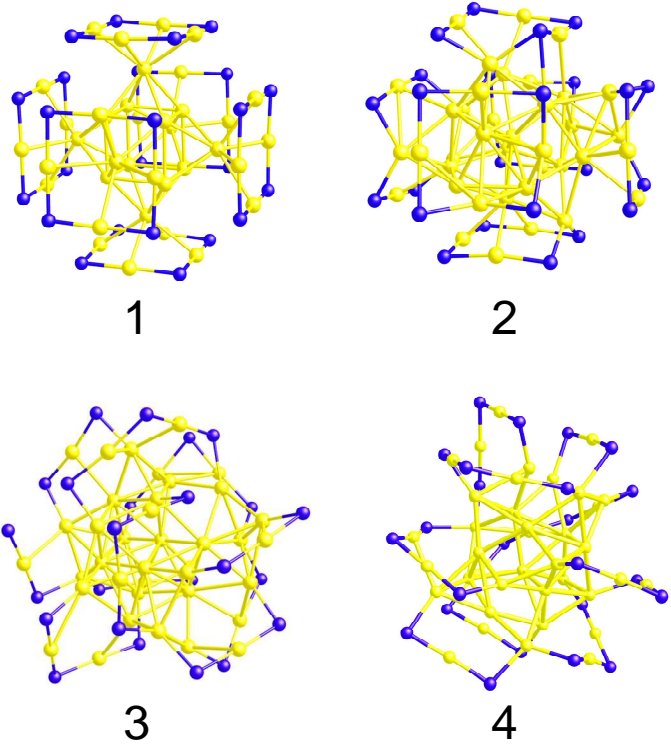

Scheme 2.

Four candidate structures for $Au_{38}(SCH_3)_{24}$. Au in yellow, S in blue, C and H not shown.

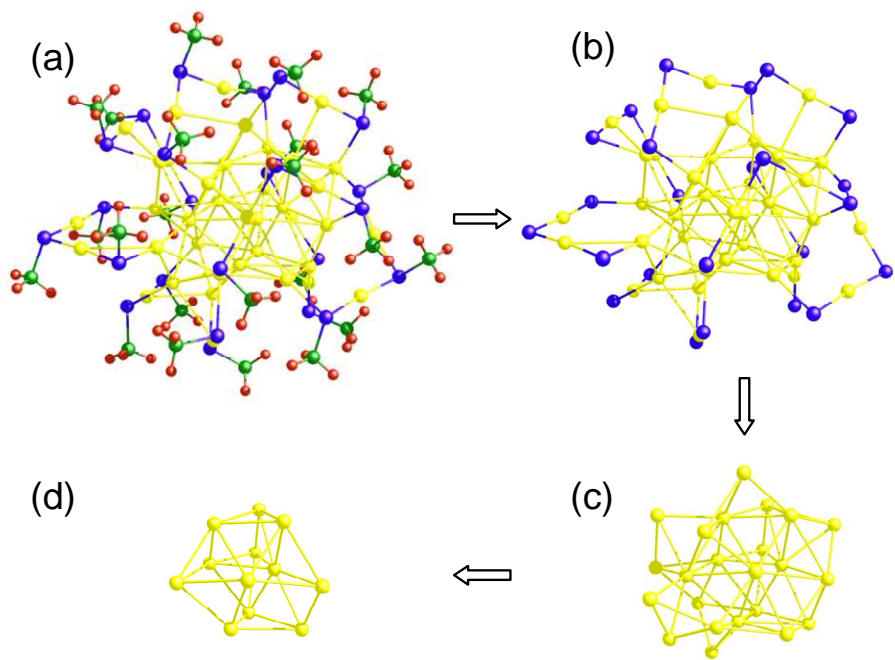

Scheme 3.

Structural details of **4**: (a) the whole cluster Au$_{38}$(SCH$_3$)$_{24}$; (b) removing CH$_3$ groups; (c) removing the staple monomers and dimers; (d) further removing Au atoms with low coordination number. Au in yellow, S in blue, C in red and H in green.

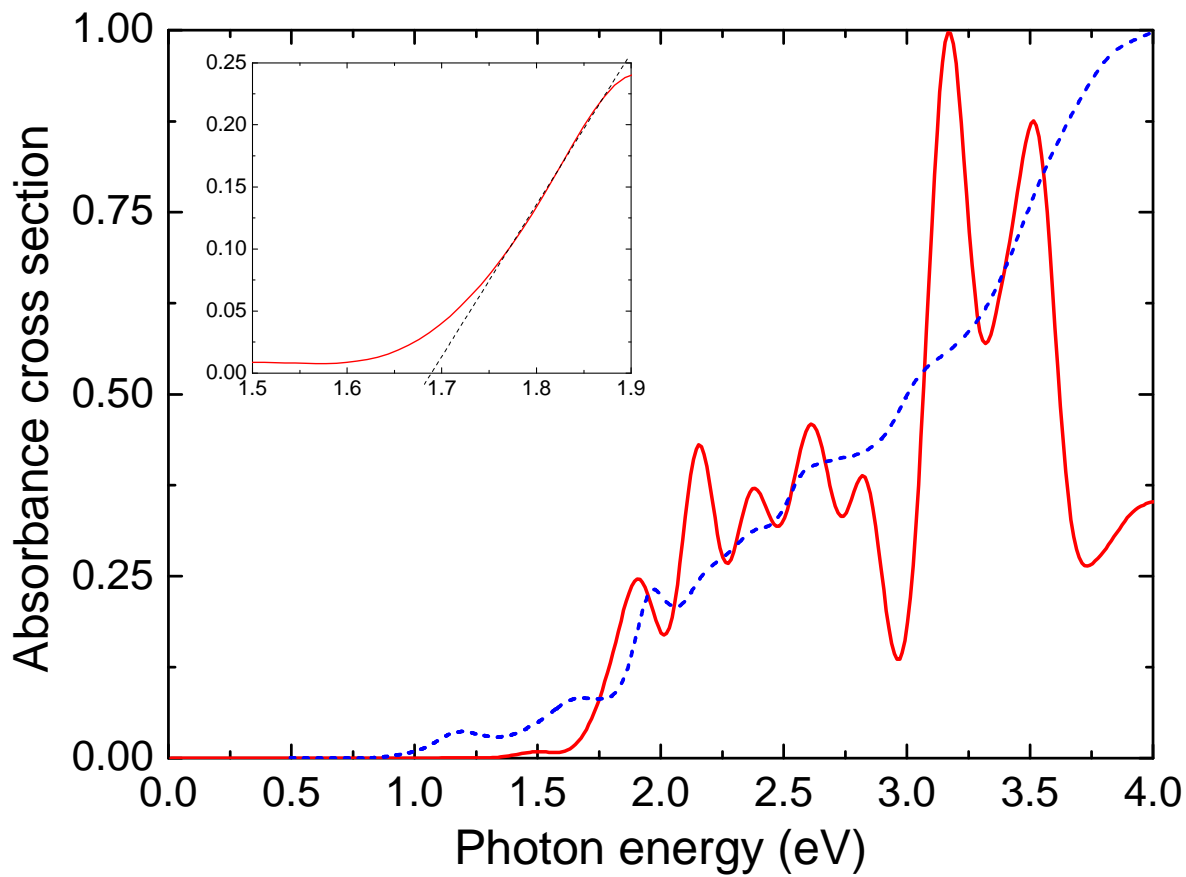

Figure 1.

Computed optical absorption spectrum for structure **1**. Experimental spectrum from Ref. 4 is shown (dashed line) for comparison. Inset shows a zoom-in of the computed spectrum between 1.5 and 1.9 eV and the extrapolation line for optical band-edge energy. Both computed and experimental spectra are normalized to strongest absorbance at or below 4.0 eV. This normalization is applied in all subsequent figures.

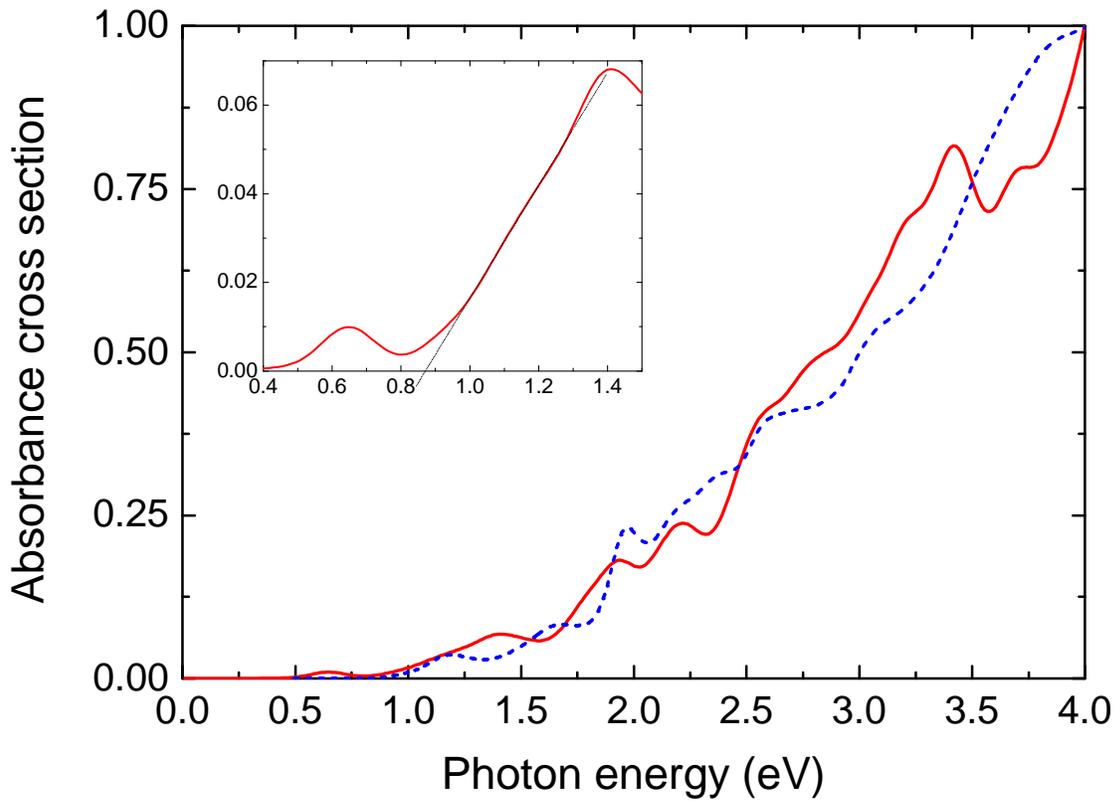

Figure 2.

Computed optical absorption spectrum for structure **2**. Experimental spectrum from Ref. 4 is shown (dashed line) for comparison. Inset shows a zoom-in of the computed spectrum between 0.4 and 1.5 eV and the extrapolation line for optical band-edge energy.

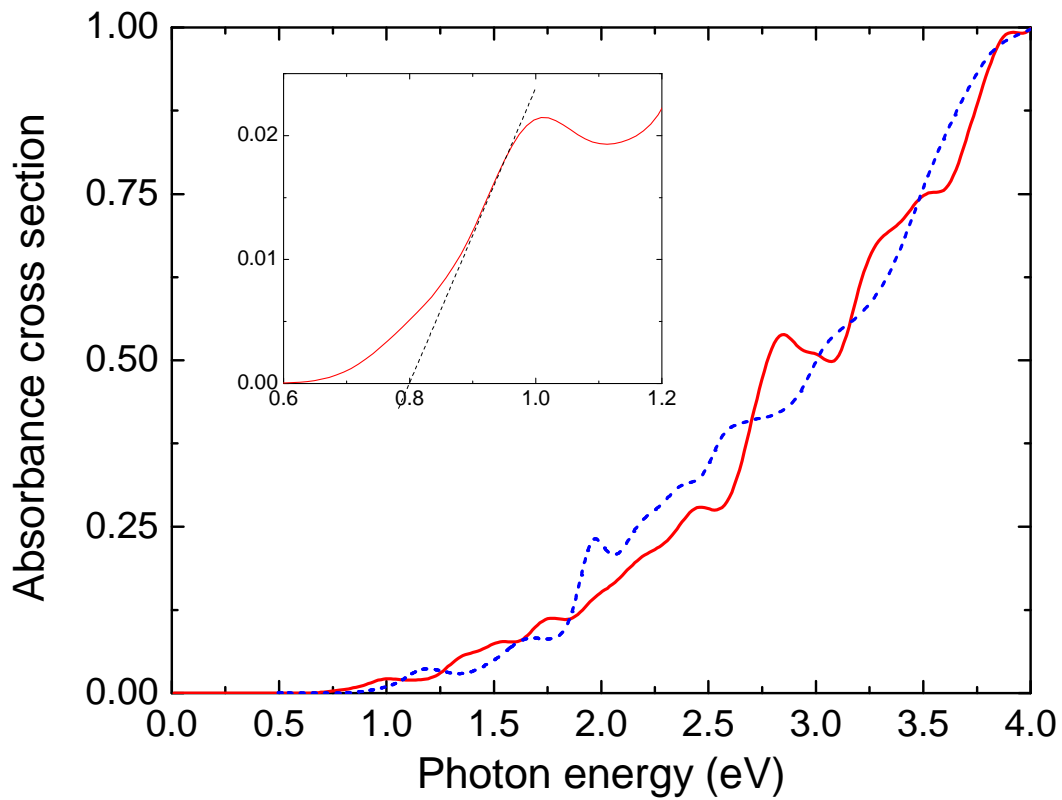

Figure 3.

Computed optical absorption spectrum for structure **3**. Experimental spectrum from Ref. 4 is shown (dashed line) for comparison. Inset shows a zoom-in of the computed spectrum between 0.6 and 1.2 eV and the extrapolation line for optical band-edge energy.

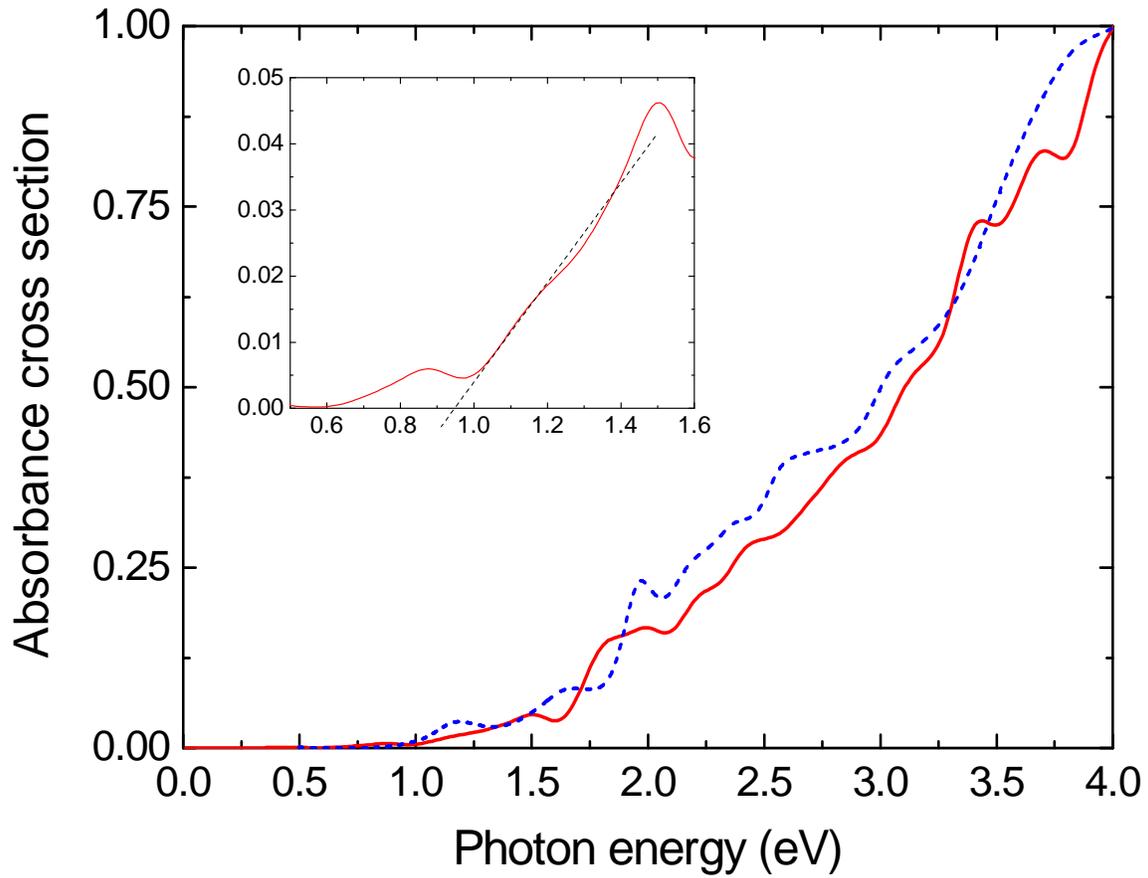

Figure 4.

Computed optical absorption spectrum for structure **4**. Experimental spectrum from Ref. 4 is shown (dashed line) for comparison. Inset shows a zoom-in of the computed spectrum between 0.5 and 1.6 eV and the extrapolation line for optical band-edge energy.

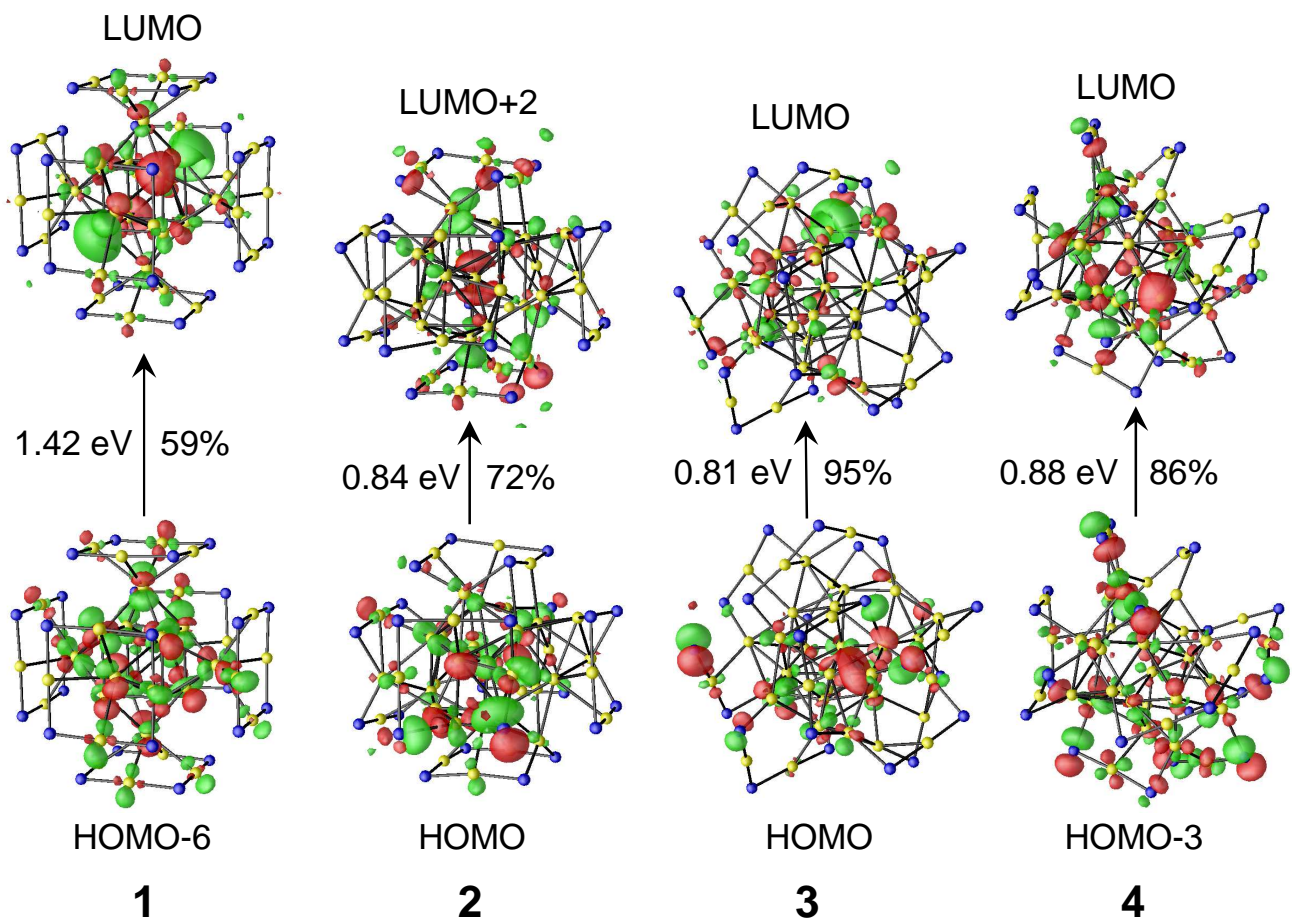

Figure 5.

Orbitals involved in single-electron transition of leading contribution to the first measurable absorption line (near the optical gap) of the computed optical absorption for structures **1**, **2**, **3**, and **4**. Line position (in eV) and percentage of leading contribution are indicated. Isovalues for green and red isosurfaces are 0.32 and -0.32 $a_0^{-3/2}$, respectively. Au in yellow, S in blue, C and H not shown.